\documentclass[12pt,a4paper]{article}

\usepackage[dvips]{graphicx}
\usepackage{amssymb}

\begin{document}
\begin{center}{\LARGE\bf Canonical DSR.}\\
\vspace{1.5em}
Franz Hinterleitner\footnote{e-mail franz@physics.muni.cz} \\
Department of Theoretical Physics and Astrophysics,\\
Masaryk University,
Kotl\'{a}\v{r}sk\'{a} 2,\\ 611\,37 Brno,
Czech Republic.
\end{center}

\begin{abstract} For a certain example of a ``doubly special relativity theory" the modified space-time Lorentz transformations are obtained from the momentum space transformations by using canonical methods. In the sequel an energy-momentum dependent space-time metric is constructed, which is essentially invariant under the modified Lorentz transformations. By associating such a metric to every Planck cell in space and to the energy-momentum contained in it, a solution of the problem of macroscopic bodies in doubly special relativity is suggested.\\[2mm]
PACS number 03.30.+p
\end{abstract}

\section{Introduction}
Preliminary quantum theories of gravity indicate a discrete quantization of space in terms of the Planck length $l_{\rm Pl}=\sqrt{\hbar G/c^3}$. This is particularly the case for Loop Quantum Gravity, which is, in fact, a quantum theory of geometry \cite{LQG}.
The hypothetical existence of a minimal length is in apparent contradiction with Lorentz symmetry in Special Relativity (SR). This has motivated a number of suggestions how to modify SR \cite{status}, \cite{Luk}, \cite{MS}. The most popular approach consists in a deformation of the action of the Lorentz group by means of a nonlinear representation in such a way that beside the vacuum velocity of light there is a second observer-independent quantity, usually an energy or a momentum. For these theories the notion Doubly Special Relativity (DSR) has become common. So far, there is no derivation of the various modifications of SR from quantum gravity, but \cite{EL} shows close relations between Loop Quantum Gravity, at least in 2+1 dimensions, and DSR.

In DSR theories the physical variables of an object are first transformed nonlinearly to so-called pseudo-variables, before the ordinary, linear Lorentz transformations act on them. Finally the inverse nonlinear transformation maps back to the physical variables in the new reference frame. There is an arbitrary number of possible nonlinear transformations, such that the Planck energy $E_{\rm Pl}=\sqrt{\hbar c^5/G}$ or the Planck momentum $p_{\rm Pl}=\sqrt{\hbar c^3/G}$ becomes invariant under the deformed Lorentz transformations, they just have to be singular at $E_{\rm Pl}$ or $p_{\rm Pl}$. So far, a few ones have been considered, called DSR 1 \cite{D1}, DSR 2 \cite{D2}, and so on.

Having formulated modified Lorentz transformations in momentum space, one may look for the corresponding transformations in space-time. This is done by invoking various guiding principles, e.\,g. \cite{Gao}, \cite{Kim}, \cite{KMS}, \cite{GR}, \cite{Mig}, normally leading to energy-momentum dependent space-time transformations. (Going the opposite way, beginning with nonlinearly modified space-time transformations, one would end up with position-dependent transformations of energy-momentum, which is hard to accept.)

The formulation in momentum space, however, leads to a problem, too. As in ordinary SR the vacuum speed of light appears as a bound on the velocity, in DSR the Planck energy or momentum appears as a bound on the energy or momentum. This makes sense as long as we are talking about ``elementary particles", but how to describe macroscopic bodies \cite{soc}? This is sometimes called the soccerball problem.  

In this paper we choose a canonical approach to construct space-time DSR-Lorentz transformations and in the sequel we construct an energy-momentum dependent space-time metric for every physical object under consideration. This suggests a concept of a local metric, such that space is partitioned into Planck cells, each with an associated metric, according to the energy-momentum content of the cell. Such a reinterpretation of $E$ and $p$ in the DSR transformations, as energy and momentum of Planck cells, can resolve the soccerball problem. 

The paper begins with a short outlay of canonical transformations in the two-dimensional enlarged phase space of a particle in section 2, in section 3 this is applied to a special form of DSR, section 4 and 5 are devoted to the construction of a space-time metric and the modification of the Lorentz contraction and in section 6 a solution of the soccerball problem is suggested.

\section{Canonical Transformations}
In the next section, starting from a transformation of momentum space variables, we are going to find prescriptions for the corresponding configuration variables, such that the complete modified Lorentz transformation, both in momentum space and in space-time, is canonical. To make things simple we are working in 2-dimensional space-time, where we consider transformations from $(t,x,E,p)$ to $(t',x',E',p')$, which are understood as the extended phase space coordinates of a physical object, typically an elementary particle, in two different inertial frames. 

For our task to find the space-time part of a canonical transformation, whose momentum space part is known, it is convenient to construct a generating function of the ``old" configuration variables $t$ and $x$ and the ``new" conjugate momenta $E'$ and $p'$. In the canonical framework we have the following relations for the partial derivatives of this generating function $F(t,x,E',p')$:
\begin{equation} \label{erz}
p=\frac{\partial F}{\partial x}, \hspace{1cm} E=-\frac{\partial F}{\partial t}, \hspace{1cm} x'=\frac{\partial F}{\partial p'}, \hspace{1cm}
t'=-\frac{\partial F}{\partial E'}.
\end{equation}
From the first two equations we determine
\begin{equation} \label{F}
F=px-Et+g(E',p'),
\end{equation}
where $E$ and $p$ have to be expressed in terms of $E'$ and $p'$ and $g$ is an arbitrary function. From the second two equations of (\ref{erz}) and eq. (\ref{F}) we obtain
\begin{equation} 
x'=\frac{\partial p}{\partial p'}\,x-\frac{\partial E}{\partial p'}\,t+\frac{\partial g}{\partial p'}, \hspace{5mm}\mbox{and}\hspace{5mm}
t'=-\frac{\partial p}{\partial E'}\,x+\frac{\partial E}{\partial E'}\,t+\frac{\partial g}{\partial E'}.
\end{equation}
The function $g$ can be determined by the following argument. Depending only on $E'$ and $p'$, it cannot depend on any Lorentz transformation. When the transformation in momentum space is unity, i. e. $E'=E$ and $p'=p$, then the space-time transformation should reduce to unity, too. For this reason $g$ must be constant, and therefore, irrelevant. We may choose $g\equiv0$, so that the final form of the generating function is
\begin{equation}
F(t,x,E',p')=p(E',p')\,x-E(E',p')\,t
\end{equation}  
and the resulting transformations are energy-momentum dependent generalizations of homogenous Lorentz transformations without translations.

\section{Application to DSR}
For reasons to be explained later the canonical program is applied to a version of DSR, which could be called a quadratic modification of DSR 2 \cite{D2}. It is defined by the transformations 
\begin{equation} \label{pseudo}
\epsilon=\frac{E}{\sqrt{1-(\lambda E)^2}} \hspace{5mm} \mbox{and} \hspace{5mm} \pi=\frac{p}{\sqrt{1-(\lambda E)^2}}
\end{equation}
to the pseudo energy $\epsilon$ and pseudo momentum $\pi$ and its inverse
\begin{equation} \label{DSR}
E=\frac{\epsilon}{\sqrt{1+(\lambda\epsilon)^2}} \hspace{5mm} \mbox{and} \hspace{5mm} p=\frac{\pi}{\sqrt{1+(\lambda\epsilon)^2}}.
\end{equation}
$\lambda$ is the inverse of the invariant energy, or equivalently, a length of the order of the Planck length. 

The quantities $E'$ and $p'$ in a different reference frame are obtained by transforming to $\epsilon$ and $\pi$, carrying out a Lorentz transformation with velocity parameter $v$ on them and transforming back to the physical energy and momentum. In the case of the theory defined above this results in 
\begin{equation} \label{Ep}
E'=\frac{\gamma E+v\gamma p}{d(E,p;v)}, \hspace{2cm} p'=\frac{v\gamma E+\gamma p}{d(E,p;v)}
\end{equation}
with
\begin{equation} \label{d}
d(E,p;v)=\sqrt{1+v\gamma^2\lambda^2(vE^2+2Ep+vp^2)}
\end{equation}
and $\gamma=1/\sqrt{1-v^2}$ as in SR.
The corresponding transformation of the space-time variables, obtained by the canonical approach outlined above, is
\begin{equation}   \label{tx}  \begin{array}{l}
t'=[(1+v\lambda^2Ep)\gamma t+(1-\lambda^2p^2)v\gamma x]\,d(E,p;v),\\[3mm]
x'=[(1-\lambda^2E^2)v\gamma t+(1+v\lambda^2Ep)\gamma x]\,d(E,p;v). \end{array}
\end{equation}
In DSR, where the transformations of $t$ and $x$ are dependent on energy and momentum, it is always possible to express the space-time transformations in terms od the ``old" momentum space variables, as above, or in terms of the ``new" ones. Equivalently to (\ref{tx}) we could express the transformations in terms of $E'$ and $p'$, inserted from (\ref{Ep}).   

It turns out that the transformed variables $t'$ and $x'$ are not orthogonal with respect to the Minkowski metric.
The coordinate axes also transform in a different way than in the case of linear Lorentz transformations. In terms of $t$ and $x$ the $t'$ axis is given by
\begin{equation}
t=-\frac{1+v\lambda^2Ep}{1-(\lambda E)^2}\frac{x}{v}
\end{equation}
and the equation of the $x'$ axis is
\begin{equation}
t=-\frac{1-(\lambda p)^2}{1+v\lambda^2Ep}\,vx.
\end{equation}
More about the nature of the transformations is revealed by an infinite boost, acting on the rest frame $(t,x)$ of an object, where $p=0$. Taking the limit $v\rightarrow1$ in this case, the $t'$ axis approaches the line
\begin{equation}
t=-\frac{1}{1-(\lambda E)^2}\,x,
\end{equation}
whereas the $x'$ axis approaches the Minkowski lightcone
\begin{equation}
t=-x,
\end{equation}
as in SR. We see that for an infinite boost the axes do not coincide.

Obviously, canonical modified Lorentz transformations and (Minkowski) orthogonality are not compatible. The linear Lorentz transformations acting in the space of pseudovariables, it would be the canonically conjugate variables to $\epsilon$ and $\pi$ that would be orthogonal with respect to the Minkowski metric, so it is not a big surprise that $t'$ and $x'$ are not. Nevertheless, for physical reasons it is desirable to have orthogonal space and time variables. If every observer should be able to identify unambiguously space and time, we have the alternative to take over the Minkowski metric  and orthogonalize $t'$ and $x'$ by a suitable linear combination $t''=\alpha t'+\beta x'$ and the corresponding $x''$ orthogonal to $t''$. In this way one would obtain pure time and space variables, which are orthogonal, but not canonical, so that the Poisson bracket $\{t'',x''\}$ would be modified. Perhaps one could rediscover a $\kappa$-deformed Minkowski algebra \cite{MS} by a certain choice of linear combinations. The great disadvantage is the arbitraryness of the choice of the orthogonal variables. In the following we shall not pursue this way but choose another alternative and construct a metric, with respect to which $t'$ and $x'$ are orthogonal. 

As a further possibility, in \cite{CH} the same realization of DSR (\ref{pseudo}) is used to construct a deformed algebra of Lorentz generators from the modified momentum space Lorentz transformations (\ref{Ep}).

\section{Construction of a space-time metric}

In SR the null directions of the metric are invariant under Lorentz transformations. Using this as a guideline for the construction of a metric in space-time, we look for the eigenvectors of the modified Lorentz transformations (\ref{tx}), written in matrix form
\begin{equation}
x'^i={\Lambda^i}_k\,x^k
\end{equation}
with the DSR-Lorentz matrix
\begin{equation} \label{Lor}
{\Lambda^i}_k=d(E,p;v)\left(\begin{array}{cc}
(1+v\lambda^2Ep)\gamma & (1-\lambda^2p^2)v\gamma \\[2mm]  (1-\lambda^2E^2)v\gamma & (1+v\lambda^2Ep)\gamma 
\end{array}\right).
\end{equation}
The function $d(E,p;v)$, introduced in (\ref{d}), is the fourth root of the determinant of the matrix (\ref{Lor}).

There are two linearly independent eigenvectors, given by multiples of
\begin{equation}
n^i_\pm=\left(\sqrt{\frac{1-(\lambda p)^2}{1-(\lambda E)^2}},\pm1\right).
\end{equation}
These invariant directions are independent of $v$, in this respect (\ref{pseudo}) was a lucky choice. (DSR 2, which (\ref{pseudo}) is a modification of, does not have this nice property.) The diagonal metric, with repect to which the vectors $n_\pm^i$ are null, it is of the form
\begin{equation}
\eta_{ik}=f\cdot\left(\begin{array}{cc} \frac{1-(\lambda E)^2}{1-(\lambda p)^2} & 0 \\[2mm] 0 & -1
\end{array}\right)
\end{equation}
with a factor $f$ yet to be determined. Constructed with the aid of eigenvectors, this form of the metric is invariant under modified Lorentz transformations, up to a change in the factor $f$. An explicit calculation, based on the invariance of the inner product of space-time vectors, $\eta_{ik}\,v^iw^k=\eta_{ik}'\,{v'}^i\,{w'}^k$, gives
\begin{equation}
\eta_{ik}'={(\Lambda^{-1})^l}_i\,{(\Lambda^{-1})^n}_k\,\eta_{ln}=\frac{1}{d(E,p;v)^4}\,\eta_{ik}.
\end{equation}
The factor $1/d^4$ is due to the fact that the determinant of (\ref{Lor}) is not equal to one.

Using the defining relations (\ref{DSR}) we confirm that the ratio betwen $\eta_{00}$ and $\eta_{11}$ is an invariant,
\begin{equation} \label{dis}
\frac{1-(\lambda E)^2}{1-(\lambda p)^2}=\frac{1}{1+\lambda^2(\epsilon^2-\pi^2)}=1-(\lambda m)^2.
\end{equation}
The last equality holds because the invariant $\epsilon^2-\pi^2$ of the linear Lorentz transformations can be expressed by the physical rest mass of the object under consideration, namely 
\begin{equation}
\epsilon^2-\pi^2=\frac{m^2}{1-(\lambda m)^2}.
\end{equation}   
(More about conservation laws and dispersion relations can be found in \cite{JV}.)
This illustrates once more the advantage of this particular version of DSR in connection with the canonical construction. With the original, more simple, DSR 2 we would not obtain a diagonal metric with this invariance property; like the eigenvectors of the Lorentz transformation, it would depend on the velocity parameter $v$.

What remains to do is to fix the factor $f$. First we recall that the modified transformations (\ref{Ep}) have the same group properties, concerning the velocity parameter, as SR: Two transformations with parameters $v$ and $w$, carried out one after the other, are equivalent to a third one with a parameter $u$ and the usual SR addition of velocities,
\begin{equation} \label{sr}
u=\frac{v+w}{1+vw}.
\end{equation}
Due to the simple dependence of $d$ on the determinant of the Lorentz transformation, in the composition of a transformation with velocity $v$ from $(E,p)$ to $(E',p')$ and a second one with velocity $w$ from $(E',p')$ to $(E'',p'')$, one has
\begin{equation} 
d(E',p';w)\cdot d(E,p;v)=d(E,p;u).
\end{equation}

We begin with the metric in the rest frame of a mass $m$ (where $E=m$ and $p=0$), where we choose $f\equiv1$,
\begin{equation} \label{rest}
\eta_{ik}^{(0)}=\left(\begin{array}{cc} 1-(\lambda m)^2 & 0 \\[2mm] 0 & -1 \end{array} \right)
\end{equation}
and go to a system, in which the velocity of the mass is given by the parameter $v$. The metric transforms to
\begin{equation}
\eta_{ik}'=\frac{1}{d(m,0;v)^4}\,\eta_{ik}^{(0)}.
\end{equation}
A further transformation with velocity $w$ results in
\begin{equation}
\eta_{ik}''=\frac{1}{d(E',p';w)^4}\frac{1}{d(m,0;v)^4}\,\eta_{ik}^{(0)}=\frac{1}{d(m,0;u)^4}\,\eta_{ik}^{(0)},
\end{equation}
with $u$ given by (\ref{sr}).
So, finally we may write the metric of a reference system, which is moving with a velocity parameter $v$ relative to a mass $m$, in dependence of $m$ and $v$,
\begin{equation}  \label{eta}
\eta_{ik}(m,v)=\left[\frac{1-v^2}{1-[1-(\lambda m)^2]v^2}\right]^2\left(\begin{array}{cc} 1-(\lambda m)^2 & 0 \\[2mm] 0 & -1 \end{array} \right).
\end{equation}

Thus, if we consider a nonvanishing rest mass $m$ and assume that $t'$ and $x'$ are its time and space coordinates in a reference frame moving with respect to this mass, then the metric, with respect to which $t'$ and $x'$ are orthogonal, depends on $m$ and the velocity.  By this dependence the metric becomes associated to a physical object. A reasonable way to make physical sense of this is to say that a massive object in the background influences causality, with the domain of influence to be discussed in the next section. 
The causality structure should be tested by the speed of massless particles or of particles very light in comparison to $m$.
Given the above metric (\ref{rest}), in the presence of a nonvanishing rest mass a vector $n^i$ is null, if $n^1=\pm\sqrt{1-(\lambda m)^2}\,n^0$. 
From this we may deduce that the velocity of light in the presence of massive objects, 
\begin{equation} \label{cm}
c_m=c\sqrt{1-(\lambda m)^2},
\end{equation}
is smaller than in vacuum.

The form (\ref{eta}) of the metric in terms of a rest mass and a velocity parameter is not the only convenient form. As $v$ has no simple physical meaning in DSR ($v$ is not directly the velocity as in SR), in some cases it may be more suitable to express the metric in terms of the object's energy and momentum. 
As mentioned below (\ref{tx}), DSR space-time transformations can always be expressed in terms of the initial energy and momentum or in terms of the final ones, the same applies for the metric. In the above form it is expressed in terms of the initial (rest frame) energy $m$ (and initial momentum zero), although in the frame, to which (\ref{eta}) applies, the object has energy $E$ and momentum $p$.
The relation between $E$, $p$ and $m$, $v$ is obtained from the transformation (\ref{Ep}) with $E=m$, $p=0$ and $E'$ and $p'$ replaced by $E$ and $p$:
\begin{equation}
E=\frac{\gamma m}{\sqrt{1+v^2\gamma^2\lambda^2m^2}}, \hspace{8mm} p=\frac{v\gamma m}{\sqrt{1+v^2\gamma^2\lambda^2m^2}} \hspace{8mm} \Rightarrow \hspace{8mm} v=\frac{p}{E}.
\end{equation}
Now we may write the metric in terms of the actual energy and momentum in a certain system,
\begin{equation} \label{epl}
\eta_{ik}(E,p)=[1-(\lambda p)^2] \left(\begin{array}{cc} 1-(\lambda E)^2 & 0 \\[2mm] 0 & -1+(\lambda p)^2 \end{array}\right).
\end{equation}

Concerning the concept of velocity in DSR there are various approaches in the literature. Whereas (\ref{cm}) gives the maximal speed of signals on the background of a mass $m$, \cite{Vel} derives the velocity of point particles on the basis of deformed Poisson brackets of its phase space variables and \cite{Gv} calculates group velocities of wave packets. Beside these two mentioned examples, \cite{Def} gives a brief overview over further different constructions of velocity. 

Here, having for disposition a space-time metric, the most natural way to define velocity is to consider tangent vectors to space-time curves. Take $w=(1,0)$ to be the tangent vector to the worldline of a mass $m$ at rest. Apply to it a Lorentz transformation with the energy-momentum parameters $E=m$, $p=0$ and the boost parameter $v$, resulting in
\begin{equation}
w'=\left( 1, [1-(\lambda m)^2]\,v\right),
\end{equation}
which we consider as the space-time velocity vector of $m$ in a boosted frame. Note that the vectors $w$ and $w'$ are not normalized to unity  with respect to the DSR metric. With the aid of the metric (\ref{rest}) we obtain the physical spatial velocity
\begin{equation}
v_{\rm ph}=\frac{\sqrt{\eta_{11}}w^1}{\sqrt{\eta_{00}}\,w^0}=\sqrt{1-(\lambda m)^2}\,v,
\end{equation}
in accordance with the expression (\ref{cm}) for the bound on velocities.

\section{A simple application: Modified Lorentz contraction}

Even if DSR theories are usually written in momentum space, the original motivation is the contradiction between continuous Lorentz contractions in space-time and the hypothetical existence of an invariant minimal length. Of course, one can simply argue that to measure a length $l$ an energy of the order of magnitude $1/l$ is necessary, and if the energy of particles is limited by the Planck energy, then the smallest length to be measured should be of the order of the Planck length. In the present theory, however, this simple argument is not always valid. For a photon, on the one hand the energy is bounded by $1/\lambda$, on the other hand, due to its zero rest mass the Minkowski metric and the Lorentz matrix are not modified.

Having a space-time metric at our disposal, we can explicitly calculate the Lorentz contraction for massive objects. Assume we have a particle with rest mass $m$ and a typical length $l$ in its rest frame. In a way completely analogous to SR we can calculate the length $l'$ in a frame moving with respect to the particle. The result is
\begin{equation} \label{contra}
l'=\sqrt{1-v^2[1-(\lambda m)^2]}\cdot l.
\end{equation}
For $v\rightarrow1$ this does not go to zero, but to the finite limit
\begin{equation} \label{lmin}
l_\infty=\lambda m\cdot l.
\end{equation}
Here one could possibly argue that to measure the length $l$ in the unprimed system, an energy of at least $\sim1/l$ is required and if we insert the rest energy $m=1/l$ into (\ref{lmin}), we would really obtain $\lambda$ as minimal length, at least in connection with massive particles.  

So far, all this is meant for a particle with sub-planckian mass. A possible meaning of these formulae in the case of a macroscopic rod will arise in the next section. 

\section{Interpretation}
In the preceding sections certain flat metrics were associated to ``physical objects", thus providing a background for small test particle dynamics. But, usually massive objects have finite spatial extension and usually there are many of them around in space. In \cite{GR} the issue of energy-momentum dependent metrics is solved in the way that every particle sees the metric of flat space-time in a different, i.\,e. energy-momentum dependent manner, so that the latter is simultaneously covered by a huge number of different global metrics, called poetically ``gravity's rainbow". Here, on the contrary, we will assume the extension of the domain of a certain metric to be bounded by the spatial extension of the material, to which it is attached. 

To carry this out we divide the space occupied by a body into (here one-dimensional) cells of a Planck volume in the rest frame of each body. This can be carried out at the scale of elementary particles as well as at the larger scale of atoms and molecules or even in the framework of mechanics of macroscopic bodies with continuous mass density. Portions of empty space between the bodies may be divided with respect to any frame, because in empty space SR is not affected. In the present framework space filled only with photons is ``empty", too, because the theory is unsensitive to massless fields and particles. Then, acording to the rest mass contained in a cell, a metric of the form (\ref{rest}) is associated to it. 

Thanks to the smallness of the Planck length for matter under normal conditions there will hardly arise the necessity to consider more than one particle in a cell. Nevertheless, the modifications of SR arise in a regime of extremely high density, where the occurrence of more particles, generally in diferent states of motion, in one cell becomes unavoidable. In this case we carry out the partition of space in the rest frame of the centre of mass of the cluster of particles which are not separable from each other by different Planck cells. Note that in DSR the physical momentum is not additive and that the total momentum, which is zero in the centre-of-mass system, is given by
\begin{equation}
p_{\rm tot}=\frac{\pi_{\rm tot}}{\sqrt{1+\lambda^2\epsilon_{\rm tot}^2}}=\frac{\sum\pi_i}{\sqrt{1+\lambda^2(\sum\epsilon_i)^2}}.
\end{equation}
In other words, in the rest frame of a composed system the sum of the pseudo-momenta is zero. 

The total mass in a Planck cell inhabited by more particles is calculated on the base of the addition formula of masses, obtained as a special case of the DSR energy composition formula 
\begin{equation}
E_{\rm tot}=\frac{\epsilon_{\rm tot}}{\sqrt{1+\lambda^2\epsilon_{\rm tot}^2}}
\end{equation}
in the special case $p_i=0$, $E_i=m_i$. For two masses this leads to
\begin{equation}
m_{\rm tot}=\frac{m_1\sqrt{1-(\lambda m_2)^2}+m_2\sqrt{1-(\lambda m_1)^2}}{\sqrt{1+2\lambda^2m_1\sqrt{1-(\lambda m_1)^2}m_2\sqrt{1-(\lambda m_2)^2}-\lambda^4m_1^2m_2^2}}.
\end{equation}

Here an input from a complete theory of gravity and its interaction with matter would be desirable. Hopefully, such a theory can decide in a natural way in which quantum of space a particle is located. For the moment we have introduced ``quanta of space" by hand by the construction of Planck cells.

From the point of view of a different observer the Planck cells may have different volumes, according to the modified Lorentz contraction, and their energy-momentum content changes according to (\ref{Ep}), but the rest mass and, in consequence, the local metric, remains unchanged. 
 
The arising picture of space is a patchwork of Planck-size cells, each endowed with a local metric, whose light-cone is the narrower, the higher the rest mass in the considered cell. At the limit of one Planck mass per Planck cell the lightcone closes and the metric becomes singular. The main consequences of such an interpretation are:
\begin{itemize}
\item With $E$ being an energy density - $E/$Planck cell - rather than the total energy of an object, the modified Lorentz transformations in momentum space provide an upper limit of the energy per Planck cell instead of the energy of a particle -- this would solve the ``soccerball problem".
\item From the spatial Lorentz transformations it follows that length contraction also depends on the local energy. With $m$ being the mass per Planck cell, (\ref{contra}) and (\ref{lmin}) describe the contraction of a macroscopic body. There is no Lorentz contraction when the mass density has the Planck value.  
\end{itemize}
In \cite{GR} flat or curved space-time as a whole is considered from the point of view of different physical objects, here we consider different objects and their attached space-time elements from the point of view of some observer with small mass density in comparison with one Planck mass per Planck volume.

In the continuum limit the different metrics in different Planck cells result in a metric field with curvature, if the mass density in some regions is sufficiently large. In effect, the obtained theory is a kind of 2-d gravity and  it might not deserve the label ``special relativistic". This is in contrast, e.\,g. to \cite{Gru}, where a $\kappa$-deformed Poisson algebra does not lead to 2-d gravity effects. 

However, to modify SR, mass (energy-momentum) densities have to be so large that significant corrections are expected only in an extrapolation of the theory far into (if not beyond) the domain of General Relativity (GR), when the theory is formulated in four dimensions. The same applies to the modified length contraction, so that these results cannot claim to have a direct physical meaning in a framework that neglects GR. As all of its effects are supposed to happen, whensoever, in the regime of very strong gravity, a theory of the kind of the one presented here seems to have a physical meaning only together with GR. A mass dependence of the metric like in (\ref{eta}) could possibly be the result of some isolated $\lambda$-dependent higher-order correction to GR or could be obtained by introducing the mass-dependent speed of light (\ref{cm}) into the coupling of gravity to matter.

As already mentioned, an analogous canonical construction for DSR 2 does not yield a comparable result. It leads to a space-time transformation with eigenvectors dependent on the boost parameter and so does not allow the construction of an observer-independent orthogonal metric. This fact led to the modification of DSR 2 used in this paper. It is not yet clear how specific the obtained result is for the chosen version of DSR, i.\,e. how restrictive the postulate of the existence of an invariant orthogonal space-time metric would be on possible realizations of DSR. The question of the space-time metric in other kinds of DSR and in other than the canonically constructed space-time representations is currently under investigation.   

In summary, the presented example leads to essentially Lorentz-invariant metrics in a modified sense and, in the case of nonzero rest masses involved, to Lorentz contractions that end up at finite lengths. The fact that this does not apply to the wavelength of photons and the weak argument that leads from (\ref{lmin}) to an invariant length (what if the length $l$ in the rest frame is measured with photons?) may be considered as unsatisfactory, but (\ref{DSR}) is only a tentative example. Its main merit is the capability to solve the problem of macroscopic bodies. \\

\noindent{\large\bf Acknowledgement.} The author thanks the Czech ministry of education for support (contract no. 143100006).

\end{document}